\documentstyle[twoside,fleqn,npb,epsfig]{article}

\title{Structure function evolution at next-to-leading order and beyond}

\author{Andreas Vogt%
\address{Instituut-Lorentz, University of Leiden, P.O. Box 9506, 
 2300 RA Leiden, The Netherlands}%
\thanks{Work supported by the EC network `QCD and Particle Structure' 
 under contract No.~FMRX--CT98--0194.}}

\begin{document}

\begin{abstract}
Results are presented of two studies addressing the scaling violations
of deep-inelastic structure functions. Factorization-scheme independent
fits to all $ep$ and $\mu p$ data on $F_2$ are performed at 
next-to-leading order (NLO), yielding $\alpha_s(M_Z) = 0.114 \,\pm\, 
0.002_{\rm exp}\, (+0.006-0.004)_{\rm th}\, $. In order to reduce the 
theoretical error dominated by the renormalization-scale dependence, 
the next-higher order (NNLO) needs to be included. For the flavour 
non-singlet sector, it is shown that available calculations provide 
sufficient information for this purpose at $x > 10^{-2} $.
\end{abstract}

\maketitle

\section{Introduction}

One of the important objectives of studying structure functions in 
deep-inelastic scattering (DIS) is a precise determination of the QCD 
scale parameter $\Lambda$ (i.e., the strong coupling $\alpha_s$) from 
their scaling violations. In this talk we briefly present results of 
two studies \cite{BV99,NV99} aiming at an improved control and a 
reduction of the corresponding theoretical uncertainties. 

\section{Flavour-singlet evolution in NLO \cite{BV99}}

The evolution of structure functions is usually studied in terms of 
scale-dependent parton densities and coefficient functions. In this 
case the predictions of perturbative QCD are affected by two unphysical 
scales: the renormalization scale $\mu_r$ and the mass-factorization 
scale $\mu_f$. While the former is unavoidable, the latter can be 
eliminated by recasting the evolution equations in terms of observables
\cite{phys}. In the flavour-singlet sector, this procedure results in 
\begin{equation}
\label{aveq1}
 \frac{d}{d \ln Q^{2}}
 \left(\!\!\begin{array}{c} F_{2}  \\ F_{B}  \end{array}\!\! \right)
 = {\cal P}\, \big(\alpha_s(\mu_r), \frac{\mu_r^2}{Q^2} \big) \otimes
 \left(\!\!\begin{array}{c} F_{2}  \\ F_{B}  \end{array}\!\!\right)
\end{equation}
with $F_B = dF_2/ d \ln Q^2$ or $F_B = F_L$. The kernels ${\cal P}$ are
combinations of splitting functions and coefficient functions which
become prohibitively complicated in Bjorken-$x$ space at NLO. Thus
Eqs.~(\ref{aveq1}) are most conveniently treated using modern complex 
Mellin-moment techniques \cite{mell}.

We have performed leading-twist NLO fits to the $F_2^p$ data of SLAC, 
BCDMS, NMC, H1, and ZEUS. Statistical and systematic errors have been 
added quadratically, the normalization uncertainties have been taken 
into account separately. The singlet/non-singlet decomposition has been 
constrained by the $F_2^n/F_2^p$ data of NMC. The initial shapes 
$F_{2,B}(x,Q_0^2)$ are expressed via standard parametrizations for 
parton densities at $\mu_f = Q_0 $.

\begin{figure}[h]
\vspace*{-8mm}
\centerline{\epsfig{file=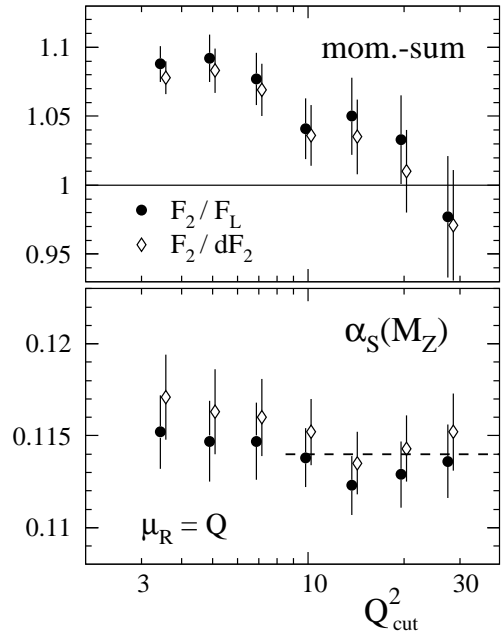,width=6.4cm,angle=0}}
\vspace*{-8mm}
\caption{The dependence of the fit results for the energy-momentum 
 sum and for $\alpha_s(M_Z)$ on the $Q^2$-cut imposed in addition
 to $W^2 \! >\! 10 \mbox{ GeV}^2$.}
\end{figure}

In order to establish the kinematic region which can be safely used for 
fits of $\alpha_s$ in the leading-twist NLO framework, the lower 
$Q^2$-cut applied to the data has been varied between 3 and 30 GeV$^2$. 
When the normalized momentum sum of the partons defining the $F_{2,B}$ 
initial distributions is left free, the fits with $Q^2_{\rm cut} \! <\! 
10 \mbox{ GeV}^2$ prefer values significantly different from unity, see 
Fig.~1. Also shown in this figure is the $Q^2_{\rm cut}$-dependence of 
the fitted values for $\alpha_s(M_Z)$, now imposing the momentum sum
rule. The results for $Q^2_{\rm cut} \leq 7 \mbox{ GeV}^2$ tend to lie 
above the $Q^2_{\rm cut} \geq 10 \mbox{ GeV}^2$ average of 
$\alpha_s(M_Z) = 0.114$ (dashed line).

In Fig.~2 we display the renormalization scale dependence of the 
$\alpha_s(M_Z)$ central values for the safe choice $Q^2_{\rm cut} = 10 
\mbox{ GeV}^2$. The conventional, but somewhat ad hoc, prescription
of estimating the theoretical error by the variation due to $0.25 \leq 
\mu_r^2\, /\, Q^2 \leq 4$ results in 
\begin{equation}
 \alpha_s(M_Z) = 0.114 \pm 0.002_{\,\rm exp} \! 
 {\footnotesize \begin{array}{c} + \, 0.006 \\ -\, 0.004 \end{array}}
 \!\Big|_{\rm scale} \: \: .
\label{aveq2}
\end{equation}
Other theoretical uncertainties are considerable smaller and can be
neglected at this point. The uncertainty due to possible higher-twist 
contributions, for instance, can be estimated at about 1\% via the 
target-mass effects included in the fits.

\begin{figure}[hbt]
\vspace*{-6mm}
\centerline{\epsfig{file=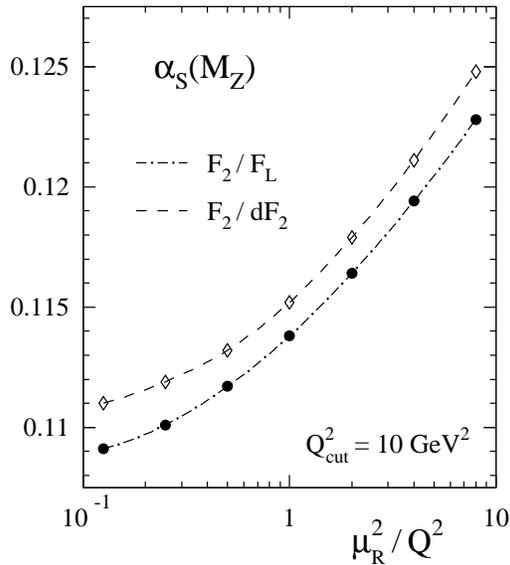,width=6.4cm,angle=0}}
\vspace*{-9mm}
\caption{The dependence of the optimal values for $\alpha_s(M_Z)$
 on the renormalization scale $\mu_r$.}
\vspace*{-1mm}
\label{avfig2}
\end{figure}

\section{Non-singlet evolution in NNLO \cite{NV99}}

The theoretical error in Eq.~(\ref{aveq2}) clearly calls for NNLO 
analyses. The necessary contributions to the $\beta$-function 
\cite{beta} and the coefficient functions \cite{coef} are known. 
However, only partial results are available for the 3-loop terms 
$P^{(2)}(x)$ in the splitting-function expansion ($a_s \equiv \alpha_s 
/ 4 \pi$)
\begin{equation}
 P = a_s P^{(0)} + a_s^2 P^{(1)} + a_s^3 P^{(2)} + \ldots \:\: .
\end{equation}
For the non-singlet part of $F_2$ considered here (NS$^+$), present
information comprises the lowest five even-integer moments \cite{pnsm}, 
the full $N_f^2$ piece \cite{pnsf}, and the most singular small-$x$ 
term \cite{pnsx}.

We have performed a systematic study of the constraints imposed on 
$P_{\rm NS}^{(2)+}(x)$ by these results. Four approximations spanning 
the current uncertainty range are shown in Fig.~3, together with their 
convolutions with a typical input shape.

\begin{figure}[htb]
\vspace*{-7mm}
\centerline{\epsfig{file=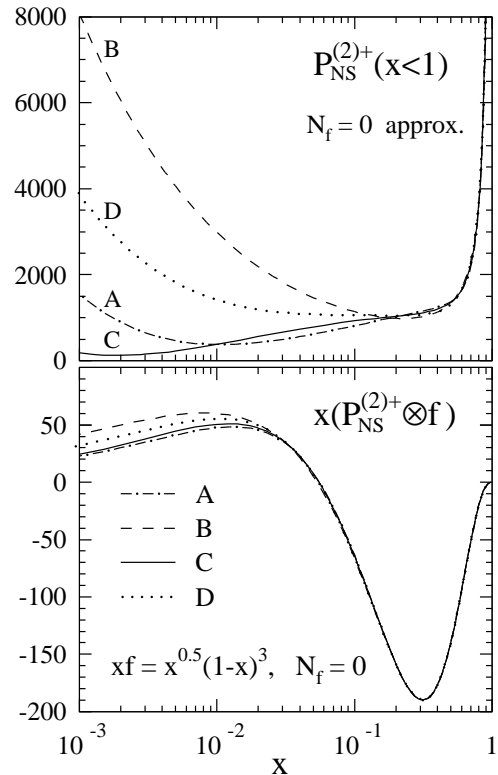,width=6.4cm,angle=0}}
\vspace*{-9mm}
\caption{Representative approximate results for the flavour-number 
 independent part of the \mbox{3-loop} non-singlet$^+$ $\overline
 {\mbox{MS}}$ splitting function.}  
\vspace*{-1mm}
\label{avfig3}
\end{figure}

\begin{figure}[htb]
\vspace*{1mm}
\centerline{\epsfig{file=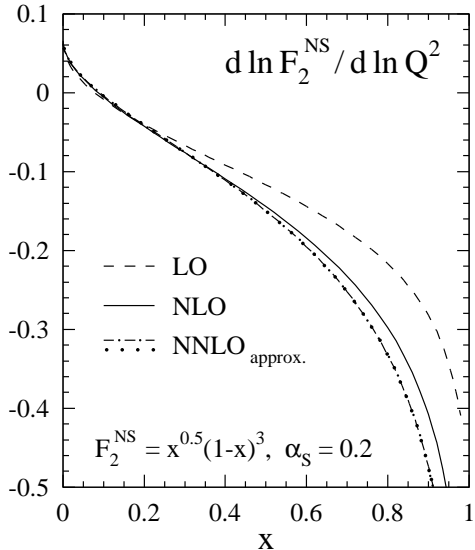,width=6.3cm,angle=0}}
\vspace*{-9mm}
\caption{The first three steps in the expansion of the scaling 
 violations of the non-singlet component of $F_2$ for typical input
 parameters.}
\vspace*{-5mm}
\label{avfig4}
\end{figure}

$P_{\rm NS}^{(2)+}(x)$ is well determined at $x \geq 0.15$, with a 
total spread of about 15\% at $x \simeq 0.3$. At (non-asymptotically) 
small $x$ its behaviour is rather unconstrained despite the known 
leading $x \!\rightarrow\! 0 $ contribution. As the splitting functions 
enter scaling violations always via convolutions 
\begin{equation}
  (P \otimes f)(x) = \int_x^1 \! dy/y\:\, P(x/y)\, f(y)
\end{equation}
with smooth initial distributions $f(x)$, the residual uncertainties 
are much reduced for observables over the full $x$-range. In the 
present case they prove to be negligible at $x > 0.02$.

The net effect of the NNLO correction is finally illustrated in Fig.~4,
where the scale-derivative of $F_2^{\rm NS}$ is shown for $\mu_r = Q$ 
and $N_f = 4$, using an $\alpha_s$-value typical for the fixed-target 
region. The inclusion of this correction into fits is expected to lead 
to a slightly lower central value for $\alpha_s$ and a considerably
reduced theoretical uncertainty.

\section{Summary and outlook}

We have analyzed present $ep/\mu p$ $F_2$-data in a 
factorization-scheme independent framework \cite{BV99}. We find that 
$Q^2,W^2\! >\! 10 \mbox{ GeV}^2$ is a safe region for leading-twist NLO 
fits of $\alpha_s$. Our central value is close to that of the standard 
pre-HERA analysis in~\cite{MV92}, but lower than the recent result 
of~\cite{MRST} using a lower $Q^2$-cut of 2 GeV$^2$. The irreducible 
renormalization-scale uncertainty turns out to be larger than expected 
from~\cite{MV92}.

We have derived approximate $x$-space expressions for the 3-loop
non-singlet splitting functions $P^{(2)}_{\rm NS}$, including error 
estimates~\cite{NV99}. This approach is complementary to, but more 
flexible than, the integer-moment procedures pursued in \cite{KPS,SY99}.
The remaining uncertainties of $P^{(2)}_{\rm NS}$ are small for the 
evolution at $x > 10^{-2}$, thus allowing for detailed NNLO analyses in 
this region. An extension to the singlet case is in preparation.

\end{document}